# An Approach to Controller Design Based on the Generalized Cloud Model

UnSun Pak, YongChol Sin, ChungJin Kwak, GyongIl Ryang

Faculty of Electronics & Automation, **Kim Il Sung** University,

Pyongyang, Democratic People's Republic of Korea

**Abstract** - In this paper, an approach to controller design based on the cloud models, without using the analog plant model is presented.

**Keywords**- Cloud, Controller Design, Stability Analysis

## 1. Introduction

Generally, there are two kinds of controller design methods; the methods with plant models [1, 2, 3, 4], and the methods without plant models [5, 6, 7, 8].

And the analysis methods of control systems are presented in previous work [5, 6, 7, 8], but the methods are based only on the normal cloud models and the analog plant models.

The analysis methods of control systems not based on the plant model was never presented before.

The approach to controller design with the normal cloud model, proposed in previous works [1~10] is nonlinear and the complexity of the calculation is very high because of the logarithmic functions, and has to be implemented by the cyclic orders in every times, because if the degree of membership of the model is 0, any calculation can't be permitted.

As the controller structure is complex and nonlinear, no structure analysis of the control system and no stability analysis of the closed-loop control system were presented.

In this paper, an approach that design controllers with the triangle cloud model which needs a little calculation and is easy to control asymmetric cloud models has been proposed.

## 2. Controller Design with the Nonlinear Triangle Cloud Model

The nonlinear triangle cloud model has the three following characteristics.

I. The expected value $Ex$, the value corresponding to the center of gravity of the model in the defined area, $(x = x_c = Ex, \mu(x_c) = 1)$, $Ex = x_c$

II. The variances $En_1, En_2$, width of the curve of the model, reflect on degree to accept as a qualitative concept.

The equation of the curve of the model is defined by three parameters $Ex, En_1, En_2$

$$\mu(x) = \begin{cases} 1 - \dfrac{x - Ex}{En_1}, & (x - Ex) \geq En_1 \quad \text{AND} \quad x < Ex \\ 1 + \dfrac{x - Ex}{En_2}, & (x - Ex) \leq En_2 \quad \text{AND} \quad x \geq Ex \\ 0, & \text{the other case} \end{cases}$$

III. The super variance- $He$, the variance corresponding to the point M of the curve, $(x = Ex + \dfrac{2}{3}En, \mu = 1/3)$, reflects on the degree of discrete of the membership cloud.

The controller with this cloud model is composed of the generator of the forward cloud, the cloud model projector, and the generator of the reverse cloud. The generator of the forward cloud, corresponding to fuzzification interface, is a block that performs division of the cloud model for input variables and makes membership clouds for each division. At first, it performs division of the cloud model, which is the same as the course of fuzzy division. Next, membership cloud function for each division should be formulated and the course of it is shown by following details.

The algorithm to generate a forward asymmetric triangle cloud is following.

Step 1: generate normal distribution random variable $x'$ that has two parameters $Ex$ to the expected value and $(En_1 + En_2)/2$ to the variance.

Step 2: generate normal distribution random variable $En_1'$ that has two parameters $En_1$ to the expected value and $He$ to the variance.

Step 3: generate normal distribution random variable $En_2'$ that has two parameters $En_2$ to the expected value and $He$ to the variance.

Step 4: calculate $\mu_A(x')$ according to following equation.

$$\mu_A(x') = \begin{cases} 1 - \dfrac{x' - Ex}{En_1'}, & (x' - Ex) \geq En_1' \quad \text{AND} \quad x' < Ex \\ 1 + \dfrac{x' - Ex}{En_2'}, & (x' - Ex) \leq En_2' \quad \text{AND} \quad x' \geq Ex \\ 0, & \text{the other case} \end{cases}$$

Step 5: define $(x', \mu(x'))$ as a drop of cloud.
Step 6: repeat step 1 to step4 until drops of number $k$ is generated.

Next, the cloud model projector has to be made up and it where assume $x$ condition cloud model when the group of rules is already known by theory of the cloud model and generate the outputs to each inputs with combine of them consist of $x$ condition cloud model and $u$ conclusion cloud model.

$x$ Condition cloud model is grouped into one degree of $x$ condition cloud model, two degrees of $x$ condition cloud model, compound degrees of $x$ condition cloud model and $n$ degrees of $x$ condition cloud model by the order of the input vector, or premise of the rule.

Assume that the rules of one rule base are following.

if $x = A_j$ then $u = C_j$, then $j = 1, 2, \cdots, m$

Then, $x$ reflects on a condition of the control rule, where is called $x$ condition cloud and $u$ reflects on an inference conclusion according to condition $x$, where is called $u$ conclusion cloud (or consequence of rule ).

In the above rule, $x$ is a variable of the cloud model and $A_j$ is a member cloud, where is $A_j = xinp(Ex_{xj}, En_{xj}, He_{xj})$, $j = 1, 2, \cdots, m$ .

And, $u$ is also a variable of the cloud model and $C_j$ is a member cloud, where is $C_j = xinp(Ex_{uj}, En_{uj}, He_{uj})$ .

For instance, the process to generate first order $x$ condition cloud is following.

Generate $\mu_A(x^0)$ for the certain input $x = x^0$ of the controller from the following equation.

$$\mu_A(x^0) = \begin{cases} 1 - \dfrac{x^0 - Ex}{En_1'}, & (x^0 - Ex) \geq En_1' \quad \text{AND} \quad x^0 < Ex \\ 1 + \dfrac{x^0 - Ex}{En_2'}, & (x^0 - Ex) \leq En_2' \quad \text{AND} \quad x^0 \geq Ex \\ 0, & \text{the other case} \end{cases}$$

Next, second order $x$ condition cloud is expressed as the multiply of the group of drops of first order $x$ condition cloud as the same case of normal cloud.

In the above equation, $En_{11}' = N(En_{11}, He)$, $En_{21}' = N(En_{21}, He)$

$En_{12}' = N(En_{12}, He)$, $En_{22}' = N(En_{22}, He)$ , where $j$ is an index of the rule of cloud control and $k$ is shown as the number of the cloud drops.

When $\mu_j^k$ obtained in this way is inputted, the group of cloud drops-, $drop(u_j^k, \mu_j^k)$ , where $k$ is shown as the number of cloud drops, the output of $u$ conclusion cloud based on first order rules, is outputted.

The process to generate $u$ conclusion cloud is following.

It follows from the variables $Ex_u$- the expected value of consequence triangle membership cloud, $En_u$- the width of it and $He_u$- the variance that determine

$$u_j^k = \begin{cases} Ex_u - (w_j^k - 1)En_u', & x < Ex_x \\ Ex_u + (w_j^k - 1)En_u', & x \geq Ex_x \end{cases}$$

$$En_u' = N(En_u, He_u)$$

In the above equation, $u$ is equal $u_j^k$ because $u$ is presented as a group of $k$ cloud drops for $j^{th}$ rule.

The generator of the reverse cloud where is a process in order to calculate parameters $Ex, En$ and $He$ of triangle membership cloud from the group of cloud drops obtained is presented as $Ex_{u_j} = mean(u_j^k)$, $u_j = Ex_u$.

Finally, the output $u$ is calculated by the weighting balance.

$$u = \frac{\sum_{j=1}^{m} w_j u_j}{\sum_{j=1}^{m} w_j},$$

Consider the asymmetric triangle cloud regulator when the plant is described by $G(s) = 167.8/(s^3 + 142 s^2 + 146 s)e^{-0.25}$.

The picture is shown the result of control by using cloud controller designed that there are three clouds- $E_1 \sim \approx E_{n_1}$, $EC_1 \sim EC_{n_2}$, $U_1 \sim U_{n_3}$, $n_1 = n_2 = n_3 = 5$, $k = 3000$,

$e \in E = [a_1, b_1]$, $ec \in EC = [a_2, b_2]$, $u \in U = [a_3, b_3]$.

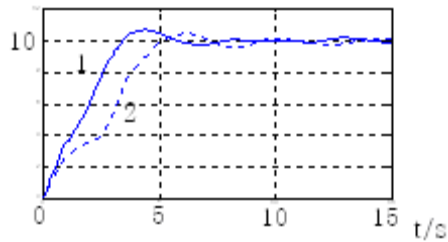

Figure1: curve that show the comparison the asymmetric triangle cloud controller with the symmetric normal cloud regulator

In the picture, first curve is shown as the control process for the asymmetric triangle cloud and its transient time is 5.1s and its steady error is 0.

Second curve is shown as the control process for the symmetric triangle cloud and its transient time is 11s and its steady error is 2.6%.

## 3. The Structure Analysis of Triangle Cloud Controller.

In order to analyze the Structure of the above cloud controller, suppose the input membership cloud functions crosses each other at their expectant value points and they are symmetric and have equal distributions.

For example, here shows a structure analysis of a 2 dimension cloud controller with 2 input variables.

Figure 2 shows the structure of the control system based on above controller.

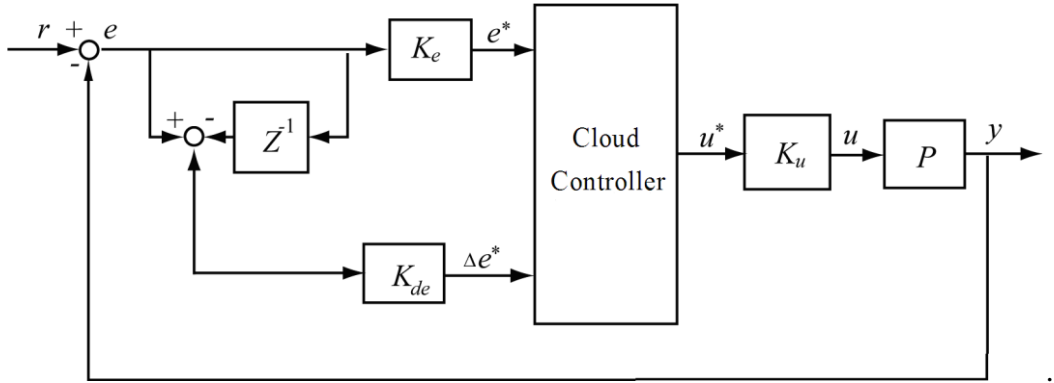

Figure 2. The structure of a control system based on 2dimension cloud controller.

**[Theorem 1]** (Structure Decomposition Expression of Triangle Cloud Controller)
A triangle cloud controller is decomposed with global nonlinear multi-value relay controller $u_G(k)$ and local nonlinear probability controller $u_L(k)$, that is,

$$K_u \Delta u^*(k) = \Delta u_G(k) + \Delta u_L(k) \tag{1}$$

(Demonstration)
Suppose that the error $e(k)$ and error derivative $\Delta e(k)$ of set value and output are the inputs of the cloud controller. And also $e^*$ and $\Delta e^*$ are the values corresponding to $e(k)$ and $\Delta e(k)$ after scaling.

$K_e$, $K_{de}$ are scaling factor of $e$ and $\Delta e$, and $K_u$ is proportionality factor, and $r$ is set value.

$$e^* = K_e e(k) = K_e(r - y(k))$$

$$\Delta e^* = K_{de} \Delta e(k) = K_{de}(e(k) - e(k-1))$$

$$e^* \in [-L, L], \quad \Delta e^* \in [-L, L]$$

$$\Delta u \in [-H, H]$$

Where $H, L$ is positive integer and usually $H = L = 1$.

Desigon of cloud controller follows below 5 steps.

① $e^*, \Delta e^*$ have $M=2J+1$ symmetric cloud variables each ; $J$ variables are positive and other $J$ variables are negative , and last is zero. (Figure 3)

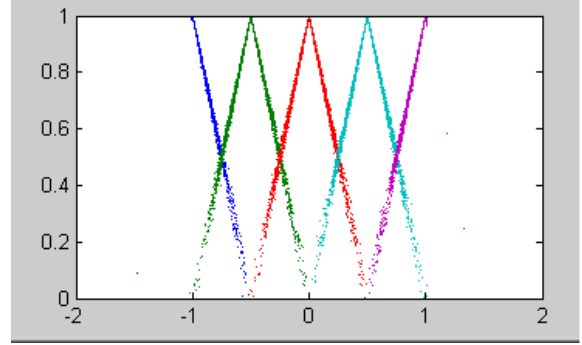

Figure 3. Symmetric premise membership cloud.

Consider the Input variable $e^*$, $\Delta e^*$ are member of triagle membership cloud set $E_i$ and

$\Delta E_j, (i, j = -J, -J+1, \cdots, -1, 0, 1, \cdots, J-1, J)$.

Each of $\mu_i$ and $\mu_j$ are the membership cloud function of $e^*$ and $\Delta e^*$, and their expectation values are $\lambda_i$ and $\lambda_j$ each.

$e^*$ and $\Delta e^*$ take certain values in below interval

$$e^* \in [\lambda_i, \lambda_{i+1}], \ \Delta e^* \in [\lambda_j, \lambda_{j+1}]$$

where

$$-J+1 \leq i, j \leq J-1$$

Then the triangle membership function width of $e^*$ and $\Delta e^*$ is as below.

$$En_e = \lambda_{i+1} - \lambda_i, \ En_{\Delta e} = \lambda_{j+1} - \lambda_j, \ = -\frac{-(i+j)}{2J} - \frac{1}{2J} + w_1 \cdot \frac{1}{2J} - w_4 \cdot \frac{1}{2J}$$

So the 2 dimension cloud model is as follows:

$$\mu_i(e^*) = (\lambda_{i+1} - e^*) / En_e', \quad En_e' = N(En_e, He_e) = \lambda_{i+1}' - \lambda_i'$$
$$= N(\lambda_{i+1}, He_e) - N(\lambda_i, He_e)$$
$$\mu_{i+1}(e) = 1 - \mu_i(e^*) = (e^* - \lambda_i) / En_e'$$
$$\mu_j(\Delta e^*) = (\lambda_{j+1} - \Delta e^*) / En_{\Delta e}', \quad En_{\Delta e}' = N(En_{\Delta e}, He_{\Delta e}) = \lambda_{j+1}' - \lambda_j'$$
$$= N(\lambda_{j+1}, He_{\Delta e}) - N(\lambda_j, He_{\Delta e})$$
$$\mu_{j+1}(\Delta e) = 1 - \mu_j(\Delta e^*) = (\Delta e^* - \lambda_j) / En_{\Delta e}'$$

where $N(\lambda_{i+1}, He)$ and $N(\lambda_i, He)$ means Gauss distribution random number that the expectation value is $\lambda_{i+1}$, $\lambda_i$ each, and their distribution is $He$.

The output variable $\Delta u^*$ has equally distributed single point membership function. (Figure 4)

There are $2M - 1$ (or $4J + 1$) cloud variables that are represented as $U_k$. Where
$$k = -2J+1, -2J, \cdots, -1, 0, 1, \cdots, 2J-1, 2J$$
$V = H/2J$, and the expectation value of $U_k$ is $kV = kH/2J$.

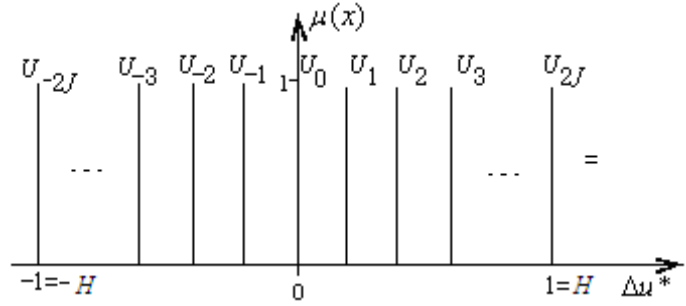

Figure 4. The Membership Cloud Function of Output variable $\Delta u^*$

② We can get $M^2$ possible combination of membership functions of $e^*$ and $\Delta e^*$. And the control rules are as follows:

IF $e^*$ is $E_i$ and $\Delta e^*$ is $\Delta E_j$ then $\Delta u^*$ is $U_{-(i+j)}$

When $e^*$ and $\Delta e^*$ are known, among the $M^2$ rules, the only 4 membership cloud functions ($\mu_i(e^*), \mu_{i+1}(e^*), \mu_j(\Delta e^*), \mu_{j+1}(\Delta e^*)$) have values and others have zero value. The below 4 rules are fired.

Rule 1: IF $e^*$ is $E_i$ and $\Delta e^*$ is $\Delta E_j$ then $\Delta u^*$ is $U_{-(i+j)}$

Rule 2: IF $e^*$ is $E_i$ and $\Delta e^*$ is $\Delta E_{j+1}$ then $\Delta u^*$ is $U_{-(i+j+1)}$

Rule 3: IF $e^*$ is $E_{i+1}$ and $\Delta e^*$ is $\Delta E_j$ then $\Delta u^*$ is $U_{-(i+j+1)}$

Rule 4: IF $e^*$ is $E_{i+1}$ and $\Delta e^*$ is $\Delta E_{j+1}$ then $\Delta u^*$ is $U_{-(i+j+2)}$

③ Design the cloud model projector.
Note the 2 dimension x condition cloud model of Rule 1~ Rule 4 as below.

Rule 1: $w'_1 = \mu_i(e^*) \cdot \mu_j(\Delta e^*) = ((\lambda_{i+1} - e^*)/En'_e)((\lambda_{j+1} - \Delta e^*)/En'_{\Delta e})$

Rule 2: $w'_2 = \mu_i(e^*) \cdot \mu_{j+1}(\Delta e^*) = ((\lambda_{i+1} - e^*)/En'_e)((\Delta e^* - \lambda_j)/En'_{\Delta e})$

Rule 3: $w'_3 = \mu_{i+1}(e^*) \cdot \mu_j(\Delta e^*) = ((e^* - \lambda_i)/En'_e)((\lambda_{j+1} - \Delta e^*)/En'_{\Delta e})$

Rule 4: $w^l_4 = \mu_{i+1}(e^*) \cdot \mu_{j+1}(\Delta e^*) = ((e^* - \lambda_i)/En'_e)((\Delta e^* - \lambda_j)/En'_{\Delta e})$

④ Generate the 1 dimension $u$ consequent cloud.

Consequent triangle membership cloud function is determined as below equations. Let's assume the single membership cloud function is used and the membership function has the expectation value $Ex_u$, and its width and distribution is 0.

$$u_1 = Ex_{u, i+j} = \frac{-(i+j)}{2J}$$

$$u_2 = Ex_{u, i+j+1} = \frac{-(i+j+1)}{2J}$$

$$u_3 = Ex_{u, i+j+1} = \frac{-(i+j+1)}{2J}$$

$$u_4 = Ex_{u, i+j+2} = \frac{-(i+j+2)}{2J}$$

As above 2 dimension $x$ condition cloud model is a group of cloud drops, we use inverse cloud projector. Then the below total output is calculated by using center method.

$$w_h = mean(w^l_h), \quad u^*(k) = \frac{\sum_{h=1}^{4} w_h u_h}{\sum_{h=1}^{4} w_h},$$

$$u^*(k) = \frac{\sum_{h=1}^{4} w_h U_h}{\sum_{h=1}^{4} w_h}$$

$$= \frac{w_1 \cdot \frac{-(i+j)}{2J} + w_2 \cdot \frac{-(i+j+1)}{2J} + w_3 \cdot \frac{-(i+j+1)}{2J} + w_4 \cdot \frac{-(i+j+2)}{2J}}{w_1 + w_2 + w_3 + w_4}$$

Here,

$$w_1 + w_2 + w_3 + w_4 = \mu_i(e^*) \cdot \mu_j(\Delta e^*) + \mu_i(e^*) \cdot \mu_{j+1}(\Delta e^*) +$$

$$+ \mu_{i+1}(e^*) \cdot \mu_j(\Delta e^*) + \mu_{i+1}(e^*) \cdot \mu_{j+1}(\Delta e^*) =$$

$$= \mu_i(e^*) \cdot \mu_j(\Delta e^*) + \mu_i(e^*) \cdot (1 - \mu_j(\Delta e^*)) +$$

$$\mu_{i+1}(e^*) \cdot \mu_j(\Delta e^*) + \mu_{i+1}(e^*) \cdot (1 - \mu_j(\Delta e^*))$$

$$= \mu_i(e^*) + \mu_{i+1}(e^*) = \mu_i(e^*) + (1 - \mu_i(e^*)) = 1$$

So the final output of cloud controller is as follows:

$$u^*(k) = w_1 \cdot \frac{-(i+j)}{2J} + w_2 \cdot \frac{-(i+j+1)}{2J} + w_3 \cdot \frac{-(i+j+1)}{2J} + w_4 \cdot \frac{-(i+j+2)}{2J} =$$

$$(w_1 + w_2 + w_3 + w_4) \cdot \frac{-(i+j)}{2J} - w_2 \cdot \frac{1}{2J} - w_3 \cdot \frac{1}{2J} - w_4 \cdot \frac{2}{2J}$$

$$= \frac{-(i+j)}{2J} - (w_1 + w_2 + w_3 + w_4) \cdot \frac{1}{2J} - w_4 \cdot \frac{1}{2J} + w_1 \cdot \frac{1}{2J}$$

$$= -\frac{-(i+j)}{2J} - \frac{1}{2J} + w_1 \cdot \frac{1}{2J} - w_4 \cdot \frac{1}{2J}$$

$$= -\frac{1}{2J}(i + j + 1 - w_1 + w_4)$$

$$= -\frac{1}{M-1}(i + j + 1) - \frac{1}{M-1}(w_4 - w_1)$$

Where

$$w_1 = \mu_i(e^*) \cdot \mu_j(\Delta e^*), \quad w_4 = (1 - \mu_i(e^*)) \cdot (1 - \mu_j(\Delta e^*))$$

and

$$w_4 - w_1 = \frac{(2e^* - \lambda_i - \lambda_{i+1})}{N(\lambda_{i+1}, He) - N(\lambda_i, He)} \cdot \frac{(2\Delta e^* - \lambda_j - \lambda_{j+1})}{N(\lambda_{j+1}, He) - N(\lambda_j, He)}$$

The output of cloud controller multipled with Proportion factor is represented as equ. (2).

$$K_u \cdot u^*(k) = -\frac{K_u}{M-1}(i + j + 1) - \frac{K_u}{M-1}(w_4 - w_1) =$$

$$= -\frac{K_u}{M-1}(i + j + 1) - \frac{K_u}{M-1}\left[\frac{(2e^* - \lambda_i - \lambda_{i+1})}{N(\lambda_{i+1}, He) - N(\lambda_i, He)} \cdot \frac{(2\Delta e^* - \lambda_j - \lambda_{j+1})}{N(\lambda_{j+1}, He) - N(\lambda_j, He)}\right] \quad (2)$$

$$= -\frac{K_u}{M-1}(i + j + 1) - \frac{K_u}{M-1}\left[\frac{(2e^* - \lambda_i - \lambda_{i+1})}{N(\lambda_{i+1}, He) - N(\lambda_i, He)} \cdot \frac{(2\Delta e^* - \lambda_j - \lambda_{j+1})}{N(\lambda_{j+1}, He) - N(\lambda_j, He)}\right]$$

As above equation, the cloud controller has 2 terms.

The first term is $-\frac{K_u}{(M-1)}(i + j + 1)$ and represented as $u_G(k)$.

Here, $e^* \geq 0$, $\Delta e^* \geq 0$.

Then note $e^* \in [\lambda_i, \lambda_{i+1}]$, $\Delta e^* \in [\lambda_j, \lambda_{j+1}]$. And $i \geq 0$ and $j \geq 0$ are const.

So $\Delta u_G(k)$ is const in the direct product space of interval $[\lambda_i, \lambda_{i+1}]$ and $[\lambda_j, \lambda_{j+1}]$. It changes when the intervals of $e^*$ and $\Delta e^*$ change.

This means $\Delta u_G(k)$ has global multi-value relay characteristics.

Figure 5 shows IO relation of $\Delta u_G(k)$.

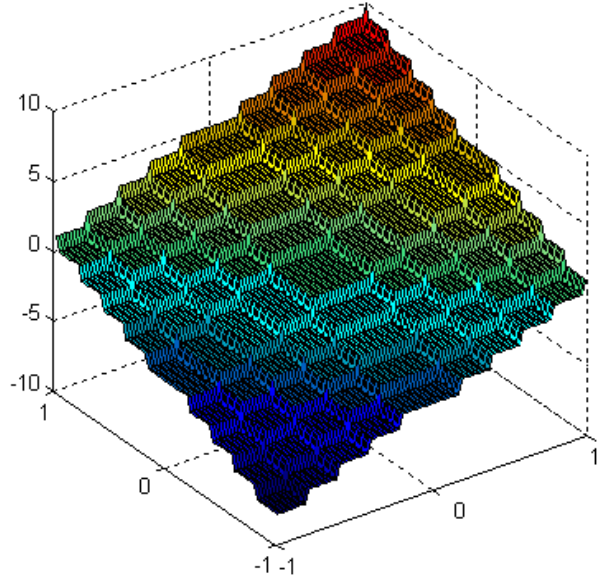

Figure 5. The IO relation of $u_G(k)$

While The second term of equ. (2) is $\Delta u_L(k)$.

$$\Delta u_L(k) = \frac{-K_u}{(M-1)}\left[\frac{(2e^* - \lambda_i - \lambda_{i+1})}{N(\lambda_{i+1}, He) - N(\lambda_i, He)} \cdot \frac{(2\Delta e^* - \lambda_j - \lambda_{j+1})}{N(\lambda_{j+1}, He) - N(\lambda_j, He)}\right] =$$

$$= K_P(e^* - \frac{\lambda_{i+1} + \lambda_i}{2}) + K_D(\Delta e^* - \frac{\lambda_{j+1} + \lambda_j}{2})$$

Where

$$K_P = -\frac{2K_u}{(M-1)(N(\lambda_{i+1}, He) - N(\lambda_i, He))},$$

$$K_D = -\frac{2K_u}{(M-1)(N(\lambda_{i+1}, He) - N(\lambda_i, He))}$$

As above equation, $\Delta u_L(k)$ is a PD controller with variable and stable statistical characteristic quantity which depends on $e^*$ and $\Delta e^*$. And its proportion factor and differential coefficient are depends on the interval and the characteristic quantity in the interval. So note that this PD controller is a local controller.

It's demonstrated in the other cases of $e^*$ and $\Delta e^*$.

## 4. Simulation

As an example of stability analysis of the cloud controller, we consider the stabilization control of a straight line inverse pendulum on a cart [11], a typical experimental instrument. The motion equation of the inverse pendulum is as follows:

$$\dot{x}_1(t) = x_2(t)$$

$$\dot{x}_2(t) = \frac{g \sin(x_1(t)) - amlx_2^2(t)\sin(2x_1(t)/2 - a\cos(x_1(t)))u(t)}{4l/3 - aml\cos^2(x_1(t))}$$

where $x_1$ is the inclined degree, $x_2$ is the angular speed and $2l$ is the length of the pendulum, $M$ is the mass of the cart, $a$ is the equivalent amplification factor, $m$ is the mass of the pendulum, $u$ is the voltage of the electric motor, $x_1 = \theta$, $x_2 = \omega = \dot{\theta}$, $g = 9.8m/s^2$, $m = 2Kg$, $M = 8Kg$, $l = 0.5m$, $a = 1/(m+M)$.

We assumed that $k_e = 0.1908$, $k_{\dot{e}} = 0.0367$, $k_u = 1.2$ and designed the cloud control system. The state equation of the closed-loop cloud control system is represented as below.

$$\begin{bmatrix} \dot{x}_1(t) \\ \dot{x}_2(t) \end{bmatrix} = \begin{bmatrix} 0 & 1 \\ 58.1 & -0.193 \end{bmatrix} \begin{bmatrix} x_1(t) \\ x_2(t) \end{bmatrix} + \begin{bmatrix} 0 \\ -0.255 \end{bmatrix} \omega(t)$$

$$y(t) = \begin{bmatrix} 1 & 0 \end{bmatrix} \begin{bmatrix} x_1(t) \\ x_2(t) \end{bmatrix}$$

When $\phi_k = 0.4$ $(k = 1, 2, 3, 4)$, $\gamma = 3$, the positive definite symmetric matrix $P_l$ s are as follows:

$$P_1 = \begin{bmatrix} 5.7838 & 0.9475 \\ 0.9475 & 8.4630 \end{bmatrix}, \quad P_2 = \begin{bmatrix} 3.5593 & 0.6051 \\ 0.6051 & 5.2080 \end{bmatrix}$$

$$P_3 = \begin{bmatrix} 4.0042 & 0.6808 \\ 0.6808 & 5.8590 \end{bmatrix}, \quad P_4 = \begin{bmatrix} 4.8940 & 0.8321 \\ 0.8321 & 7.1610 \end{bmatrix}$$

Note that the designed cloud control system is asymptotically stable.

We compared the proposed control method with LQ control method and gauss cloud control method. The designed LQ controller 's parameters are as following.

$$R = 0.1, \quad Q = \begin{bmatrix} 20 & 0 \\ 0 & 0.1 \end{bmatrix}$$

We compared the methods in the two conditions – frictionless condition and frictional condition.

Figure 8 shows the simulation result in the frictionless condition.

The initial condition is $\theta = 20$.

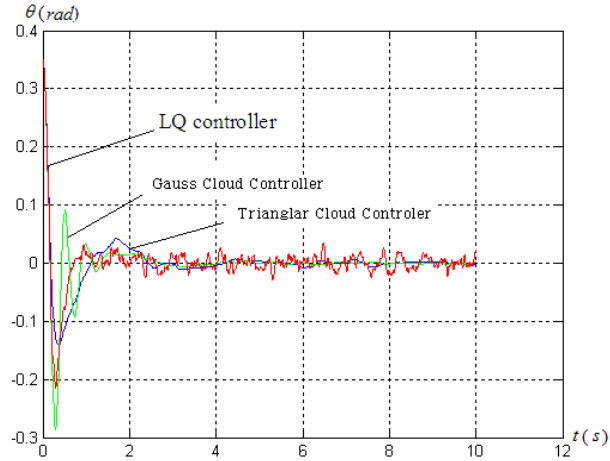

Figure 8. The simulation result of three methods in the frictionless condition.

Note that the methods are nearly similar in the frictionless condition.

Figure 9 shows the result on the condition with dry friction and viscous friction.

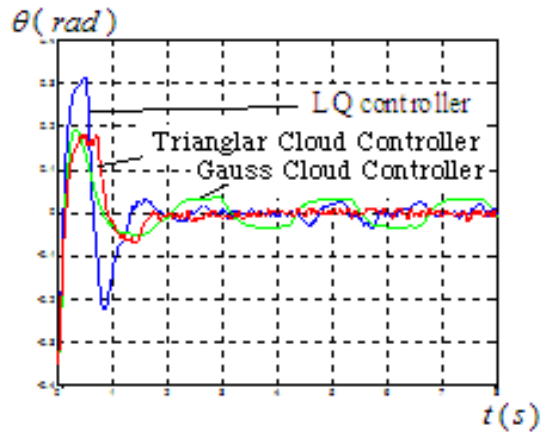

Figure 9. The simulation result of three methods in the frictional condition.

In this case, the maximum chatter width of the pendulum are 4°, 3° and 1° each. And the maximum amplitude of the pendulum are 11°, 18° and 10° each.

Now, note that the cloud controller is superior than others.

We analyzed the stability of the cloud control system by the cloud Lyapunov function and found that the cloud control system is asymptotically stable and have the cloud $H_\infty$ performance index when $k_e = 0.1908$, $k_{\dot{e}} = 0.0367$, $k_u = 1.2$.

As we presented in previous sections, the cloud controller is capable for the stabilization of nonlinear affine systems.

The cloud controller represents the several talents' functionality while the fuzzy controller represents a talent's functionality.

The shape of the membership cloud is determined by the numbers of the cloud drops. The number of the drops is in the range from 1000 to 3000.

If the number is very small, for example 100, the membership clouds may not be inclined at all. And if the number is very bigger than the ranged value, it may increase the complexity of computation while it is useless for the control performance.

And with small numbers of cloud drops, the design advantage of the cloud controller, the combination of fuzziness and probability may not displayed.

## 5. Conclusion

I. Introduced a design method of generalized asymmetric triangle cloud controller without using analog plant model.
II. Analyzed the structure of the generalized cloud controller without using analog plant model.
III. Simulation of above method proved the validation and effectiveness.

**References**

[1] Choe Song , A message in the Academy of the Science, 2011. No 6. 34-38
[2] 姜长生，高键,《一种新的云模型控制器设计》v.34，N.2, 157-162(2005)
[3] 张飞舟，范跃祖，李德毅.《基于隶属云发生器的智能控制》[J] 航空学报，v.20, N.1, 89-92(1999)
[4] 吕辉军，王晔，李德毅.《逆向云在定性评价中的应用》[J].计算机学报，v.26, N.8, 1009－1014(2003)
[5] 刘常星，等,"基于云X信息的逆向云新算法", 系统仿真学报, v.16, N.11, 2417-2420(2004)
[6] 姜长生，李众,高键,《一种新的云模型控制器设计》v.34，N.2, 157-162(2005)
[7] 李众，杨一栋，《一种新的基于二维云模型不确定性推理的智能控制器》2005,8
[8] 王洪利,冯玉强.《基于云模型具有语言评价信息的多属性群决策研究》(哈尔滨工业大学管理学院,黑龙江哈尔滨 2005 ,6
[9] 张飞舟,晏磊,范跃祖,孙先仿.《基于云模型的车辆定位导航系统模糊评测研究》 控制与决策6 2002,9
[10] 杨斌.李 忘 "一种关系数据库中基于云模型关联规则的提取》,上海交通大学学报, v.37，N.4，35-39(2003)